\documentclass[]{aa} 
\usepackage{psfig} 
\usepackage{natbib} 
\usepackage{amssymb} 
 
\begin{document}         
 
\title{Homogeneous age dating of 55 Galactic globular clusters.} 
\subtitle{Clues to the Galaxy formation mechanisms}

\author{M.~Salaris\inst{1,2} \and A.~Weiss\inst{2}} 
 
\institute{Astrophysics Research Institute, Liverpool John Moores 
University, Twelve Quays House, Egerton Wharf, Birkenhead CH41 1LD, UK 
\and 
Max-Planck-Institut f\"ur Astrophysik, 
Karl-Schwarzschild-Str.~1, 85748 Garching, 
Federal Republic of Germany} 
 
\offprints{M.~Salaris; (e-mail: ms@astro.livjm.ac.uk)} 
\mail{M.~Salaris} 
 
\date{Received; accepted} 
 
\authorrunning{M.~Salaris \& A.~Weiss} 
\titlerunning{Ages of a large sample of Galactic globular clusters}

\abstract{We present homogeneous age determinations for 
a large sample of 55 Galactic 
globular clusters, which constitute about 30\% of the total Galactic 
population. A study of their age distribution  
reveals that all clusters from the most metal poor ones up to intermediate 
metallicities are coeval, whereas at higher [Fe/H] an age spread exists, 
together with an age-metallicity relationship. At the same time, all 
clusters within a certain galactocentric distance appear coeval, whereas an 
age spread is present further away from the Galactic centre,  
without any correlation with distance. 
The precise value of [Fe/H] and galactocentric distance for the onset 
of the age spread and the slope of the age-metallicity relationship 
are strongly affected by the as yet uncertain [Fe/H] scale. 
We discuss how differences in the adopted [Fe/H] scale and cluster sample 
size may explain discrepant results about the clusters age distribution  
reached by different authors. 
Taking advantage of the large number of objects included in our 
sample, we also tested the possibility  
that age is the global second parameter which determines 
the Horizontal Branch morphology, and found indications that  
age could explain the global behaviour of the second parameter effect.
\keywords{Galaxy: formation -- Galaxy: halo -- globular clusters: general 
-- stars: Hertzsprung-Russell diagram -- stars: Population II} 
} 
\maketitle

\section{Introduction} 
 
Less than a decade ago the age of the oldest globular clusters 
appeared to be much higher than that of the expanding universe. But at 
the end of the last millenium significant improvements both in models 
and in observational data, notably in the determination of cluster 
distances by virtue of {\sl Hipparcos}-based distances, lead to a reduction 
of cluster ages. Presently, most determinations scatter around a 
typical age of the oldest objects of 12--14~Gyr 
\citep{sw:98,cgc:2000,v:00,chabp:2001}.  With the growing confidence 
in the absolute age determinations and an increasing number of 
extensive homogeneous and high-quality photometric cluster data, the 
interest has shifted to questions concerning relative ages in order to 
learn about the formation of the galaxy and its halo and disk 
components. 
 
In our own work \citep{sdw:97,sw:97,sw:98} we have used a method (to 
be fully described in Sect.~2), which is a mixture of absolute  
and relative age determinations in four metallicity ranges 
(see also \citealt{rhf:96} for a similar technique).  
It does not need the knowledge of cluster distances,  
but instead predicts them. 
To determine the absolute ages of a sample of reference clusters 
for the various metallicity ranges, 
we use the brightness difference between horizontal  
branch (HB) and turn off (TO); the HB sets the cluster 
distances and the TO (whose brightness is the best  
predicted quantity to be used as age indicator) their age.  
For relative ages with respect to reference clusters 
we consider the differences in the extension of the subgiant branch,  
i.e., the colour range between the TO and the base of the 
red giant branch (RGB). In this way we are able to determine ages also 
in case of clusters for which the HB brightness determination is  
problematic (e.g., blue and/or sparsely populated HB). Since  
the differential properties of the TO-RGB colour difference across  
the entire cluster metallicity range is in principle 
subject to current uncertainties 
in convection treatment and colour transformations, the method works 
best (i.e., the uncertainties are minimized) if restricted to  
small [Fe/H] ranges. 
When selecting the limits of the individual [Fe/H] intervals, one   
must strike a balance between having  
a sizable number of clusters, and at the same time  
covering a [Fe/H] interval where  
the predicted variation of the derivative of the TO-RGB colour 
difference with respect to age and [Fe/H] is small. 
 
In \cite{sw:98} we applied this method to a sample of 31 clusters taken from 
various sources, such that we could start to address the question of 
age-metallicity and age-galactocentric distance relations. 
This cluster sample appeared still too small  
to draw statistically meaningful conclusions; 
strictly speaking, relative ages are valid only within each 
metallicity bin, but we have shown in \cite{sw:98} and \citet{psw:98}  
that they are consistent all over the [Fe/H] range spanned by the clusters. 
In the present paper, we want to derive ages for a much 
larger sample of globular clusters by using our own isochrones and 
applying our preferred method. Relative ages for these clusters are a 
straightforward by-product. 
 
There are other age determination techniques in use, some of them being 
easier to apply, but probably less accurate. All, of course, depend at 
some stage on theoretical isochrones.  
Semi-empirical methods \citep{bcpf:98,rspa:99} are 
of particular interest for relative age-dating; they make extensive 
use of the colour difference between TO and RGB across all the [Fe/H] 
range spanned by the galactic globular clusters,  
thus avoiding the necessity of RR~Lyrae 
stars or a well-developed HB for distance 
determination. This allows to apply  
the method to a much larger number of clusters, and thus to obtain a 
solid picture of cluster formation throughout the Galactic history.  
The calibration of the TO-RGB colour difference with 
respect to age and metallicity is obtained empirically making use of 
clusters of different ages and [Fe/H], for which reliable 
ages from the TO-HB brightness difference can be determined. 
\cite{bcpf:98} noticed inconsistencies in relative 
cluster ages derived from brightness differences (between HB and TO) 
or colour differences as predicted by the isochrones they employed. In 
\cite{psw:98} we could demonstrate the self-consistency of our own 
isochrones for $(B-V)$-colours. 
 
Following on this, \cite{rspa:99} applied the semi-empirical method 
for colour differences in $(V-I)$ to their own large sample of cluster 
data \citep{rasp:2000,rpsa:2000}. Their main result, based on the 
isochrones by \cite{scl:97} and a pre-release of \cite{vsria:2000}  
models, was 
that clusters of $\mathrm{[Fe/H]} \le -1.2$ are probably coeval 
(within 1 Gyr) and that at higher metallicities, up to 
$\mathrm{[Fe/H]} \le -1.0$ a few clusters younger by $\approx 4$~Gyr 
exist, too, and finally, that at $\mathrm{[Fe/H]} \approx -0.7$ younger 
clusters prevail. However, if the brightness difference between HB  
(whose absolute magnitude  
was taken from the empirical relation by \citealt{cgc:2000})  
and TO was used as the age indicator, this picture was not reproduced 
completely (some small differences in the age distribution for the 
most metal poor and most metal rich clusters), 
and, most importantly, the absolute age calibration for 
both relative age scales differed by up to 3~Gyr. A further set of 
isochrones tested \citep{ccd:98} showed a large age spread for the 
colour-difference method. Evidently, semi-empirical methods for 
relative cluster ages spanning the whole metallicity range reveal 
different and internally inconsistent results for different isochrone 
sets, at least when one does not consider the HB brightness from ZAHB models 
consistent with the employed isochrones.  
 
In this paper we will apply all these methods  
to a cluster sample which is a 
combination of the \cite{rasp:2000,rpsa:2000} set (35 clusters) and 20 
additional clusters with accurate photometry obtained 
recently. For this, we need colour transformations to $(V-I)$; in 
\cite{ws:99} we investigated the properties of our own choice of 
transformations, showing that for selected clusters with photometry in 
$V$, $B$, and $I$ we can obtain identical ages and similar isochrone 
fitting quality for $VI$ and $BV$ photometric data. In the present 
paper we will determine cluster ages from $VI$-data using this and a 
second set of transformations \citep{aam:99}. We will demonstrate that 
the semi-empirical relative ages across the whole cluster metallicity range 
we obtain from the TO-RGB colour difference 
depend strongly on the chosen transformations; 
those by \cite{aam:99} provide results consistent with the 
ages inferred from the TO-HB brightness differences and the method 
employed in \cite{sw:98}, while the transformations used in 
\cite{ws:99} are not suited for this purpose. 
 
On the other hand, as will become evident, our preferred age 
determination method, as used by \cite{sw:98}, turns out to be largely 
insensitive to the specific colour transformation adopted, although it does 
make use of the colour difference between TO and RGB within restricted 
metallicity ranges.  
Similarly, the use of either $VI$ or $BV$-data does not 
influence the picture of the absolute and relative ages, but may affect 
only some individual clusters.  
Quantitatively, the effect is 
comparable to that of using different photometric sources; it is 
important to realize that in spite of the large observational progress 
and effort, the source of photometric data still influences individual 
and global results on cluster ages.  
 
The paper is organized as follows: In the next section, our 
preferred age determination method and its application to a large 
sample of 55 globular clusters will be presented. This is,  
to our knowledge, the 
largest sample of clusters ever investigated, using the same  
age-dating method and theoretical isochrones. The sample size 
is a significant increase over the 43 clusters 
by \citet{cds:96}; analysis of large clusters sample have been 
recently performed also by \citet[ 36 objects]{rhf:96}, 
\citet[ 33 objects]{bcpf:98} and  
\citet[ 35 objects]{rspa:99}. 
Our sample is largely based on homogeneous 
and very recent photometries; the isochrones 
and colour transformations are of the latest generation as well. 
In Sect.~3 we will then turn to the semi-empirical relative-age 
method using $(V-I)$-colours. We will present completely consistent 
results that agree well with those of Sect.~2. The discussion 
in Sect.~4 will analyze the age distribution of our cluster sample 
in light of the proposed scenarios of Galaxy formation. 
We will present also an analysis of 
the effect of the sample size and the adopted 
clusters metallicities on the derived age distribution.  
About the latter point, we have considered cluster [Fe/H] values 
derived by both \citet[ our reference metallicities]{cg:97}  
and \citet{zw:84}. The two sets of 
values show differences up to $\sim$0.3 dex (\citealt{cg:97} [Fe/H] 
being higher) for intermediate metallicity clusters; a 
quadratic relationship (\citealt{cg:97}) transforms 
\citet{zw:84} metallicities into the corresponding \citet{cg:97} ones.  
The possibility that age is the global second parameter responsible 
(together with [Fe/H]) of the cluster HB morphology will  
be also tested. A short summary of the main results follows 
in Sect.~5.

\section[]{The cluster sample and age-dating method} 
 
Our GC sample consists of 55 objects. The $V$ magnitude difference 
between Zero Age HB (ZAHB) and TO, $\Delta V$, and 
the $(V-I)$ colour difference ($VI$ Johnson-Cousins bands) 
between TO and the base of the RGB, $\Delta(V-I)$ as defined 
in \citet[ R99]{rspa:99}, for 35 of them 
are taken from the homogeneous database by 
R99. For the remaining 20 objects $\Delta V$, $\Delta (V-I)$ or  
$\Delta (B-V)$, the corresponding colour difference between TO and base of the 
RGB in $(B-V)$ as defined by \citet{vbs:90}, 
are taken from 
\citet[ SW98]{sw:98} and other papers published  
in the last 5 years (see Table~\ref{data}). 
 
The age-dating method is analogous to the 
procedure outlined in \citet[SW97]{sw:97} and SW98. 
The GC sample is divided into four metallicity 
intervals, which are the same ones as in SW98, apart from the last interval;  
this we extended to include in the same group the thick disk clusters 
present in the sample. The metallicity intervals therefore are 
$-2.15 \leq $[Fe/H]$ \leq -$1.75,  $-1.74 \leq $[Fe/H]$ \leq -$1.3,  
$-1.29 \leq $[Fe/H]$ \leq -$0.9, and $-0.89 \leq $[Fe/H]$ \leq -$0.6. 
We used the [Fe/H] values given by \citet{rhs:97} 
on the \citet[ CG97]{cg:97} scale (the internal accuracy of these 
[Fe/H] values is of the order of 0.10 dex); this introduces small differences 
with respect to the [Fe/H] values used by SW98 which were either directly 
from CG97 or transformed from the \citet[ ZW84]{zw:84} scale using the  
relationship given by CG97, but allows us to directly compare our 
results with R99, who selected the same [Fe/H] values for their sample.  
In case of clusters without estimates from \citet{rhs:97}, but with 
estimates on the ZW84 scale, we again 
transformed their [Fe/H] onto the CG97 scale  
using the conversion formula given by CG97.  
 
Within each interval a reference cluster  
showing a HB well populated  
at the instability strip region and/or at its red side 
was selected; 
this allows a reliable determination of the ZAHB level to be obtained, 
and the absolute age to be 
determined accurately from $\Delta V$, by comparing it with $\Delta V$ 
taken from the isochrone for the metallicity under 
consideration.  
The reference clusters are,  
respectively, M~15 (NGC~7078), M~3 (NGC~5272), NGC~6171 and  
47~Tuc (NGC~104). In our new analysis M~15 and M~3 have replaced 
M~68 (NGC~4590) and NGC~6584 as reference clusters in their  
respective groups (SW98).  
Due to the small [Fe/H] differences with respect to SW98, the cluster 
NGC~6171 has now moved to the third metallicity group where it has 
replaced M~5 (NGC~5904) as the reference cluster. 
For all the four reference clusters we used 
$\Delta V$ values taken from R99.  
 
Once the absolute ages of the  
reference clusters were obtained, the 
age of all other clusters within each group was determined 
differentially with respect  
to the reference one, by using the differences in $\Delta (V-I)$ 
to derive differences in age, again by using the isochrones. 
In case of clusters with $BV$ photometry, their relative age 
with respect to the template clusters was obtained from  
$\Delta (B-V)$. 
Because of the approach of tying relative ages within limited 
{\em m}etallicity ranges to the {\em a}bsolute age of selected 
reference clusters, we will, for the remainder of this paper, call our 
method simply the {\em AM}-method. 
 
We have employed in our analysis the $BV$ and $VI$ isochrones described  
in SW98 and \citet{ws:99}, assuming a cosmological helium mass 
fraction Y=0.230, an average $\alpha$-element overabundance  
[$\alpha$/Fe]=0.4, and a helium enrichement factor $\Delta$Y/$\Delta$Z=3. 
For the $(V-I)$-colours we use a combination of the 
transformations by \citet{bcp:98} for the main sequence and \citet{vb:98} 
for the RGB, as described in \citet{ws:99}. For further details, see 
the original papers (SW98, \citealt{ws:99}). We briefly recall here that 
the ZAHB luminosities from our models, which set the zero 
point of our absolute ages, are in agreement with the ZAHB 
brightness levels obtained by \citet{cgc:2000} 
from the subdwarf fitting method applied to a sample of Galactic 
globular clusters. They are  
therefore about 0.13 mag brighter than the relationship used by R99, 
which results from an average of the subdwarf fitting distances 
together with other distance determinations, discussed  
in \citet{cgc:2000}. 
 
Our procedure allows a reliable determination of the age distribution 
with respect to the absolute age of the template clusters, largely 
unaffected by uncertainties in the convection treatment, by the precise 
value of the cluster metallicity, and by the inclusion of 
element diffusion (see the discussion in SW97). It is also fairly 
independent of current uncertainties in colour transformations. This 
we verified by repeating the procedure for the clusters with $VI$ 
photometry using a different set of RGB colour transformations 
together with our isochrones (\citealt{aam:99}); we find only modest 
age changes within $\pm$0.6 Gyr. 
A more detailed discussion on the issue of colour transformations will 
be presented in the next section. 
 
\begin{figure}[ht] 
\psfig{figure=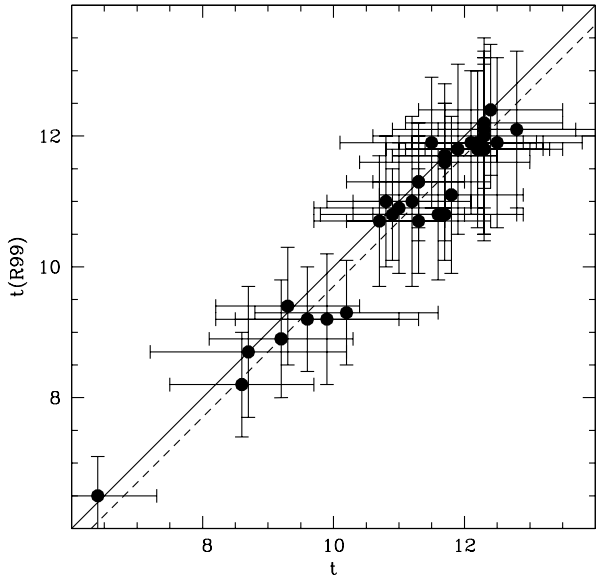,width=8.3cm,clip=} 
\caption[]{Comparison of the absolute ages (in Gyr) obtained using the  
$\Delta V$ values provided by R99 for their whole sample 
(vertical axis), with the ages obtained with our AM-method (horizontal 
axis). Errors are as shown in Table~\ref{data}, or as given by R99.
The solid line represents the locus of equal ages; dashed line 
corresponds to the average difference between the two sets of ages (0.3 Gyr).} 
\label{compage2} 
\end{figure} 
 
Two further tests have been performed in order to assess the reliability 
of our age determination: In the first one we have simply moved 
clusters close to the boundary of two contiguous groups  
(e.g., NGC~6254, NGC~3201, NGC~6397) to the 
adjacent one and redetermined their age with respect to the absolute 
age of the new template cluster. The resulting absolute ages agree 
within 0.3 Gyr with the ones obtained from the original group location. 
In the second test, we have derived the age of all 35 clusters 
from R99 making use of their estimated $\Delta V$-values for the whole sample. 
Even in case of blue HB morphologies, R99 have derived the level of the ZAHB 
extrapolating the observed ZAHB location towards the red, by fitting   
template cluster HBs uniformly populated to the observed 
branches. While we consider this approach being less accurate then in cases 
with observed HBs populated at the instability strip or at its red side,  
it is of course a valid estimate of the ZAHB brightness.  
The resulting ages for the 35 clusters, displayed in Fig.~\ref{compage2},  
indeed agree extremely well with the  
values obtained with the AM-technique, with an average difference of 
only 0.3 Gyr ($\Delta V$ ages being younger). 
This test confirms the  
mutual consistency of the $\Delta V$ relative ages and those obtained 
by the AM-method over the whole [Fe/H] range spanned by the cluster sample. 
 
A similar test, but in $(B-V)$, was performed for two individual 
clusters, NGC~6712 and  
NGC~6934. $BV$ photometry is available for both of them, and both have 
extremely well-defined ZAHB levels. We could therefore derive their 
absolute age also directly from $\Delta V$. As in the case of the R99 cluster 
sample with $VI$ photometry, the results are consistent, within 0.4 Gyr, 
with the relative age obtained from their $\Delta (B-V)$ with 
respect to the corresponding template cluster.

\begin{table*} 
\renewcommand{\arraystretch}{0.95} 
\caption[]{Data for the 55 clusters analyzed. The columns display, 
respectively, cluster name, age using the CG97 scale and associated error, [Fe/H] on the CG97 scale, 
age using the ZW84 scale and associated error, [Fe/H] on the ZW84 scale, galactocentric distance, HB type, 
photometric bands used for the age determination,  
source of the data (see text for details). Reference clusters are 
emphasized by bold type characters.} 
\begin{minipage}{\textwidth} 
\begin{tabular}{lrcrcrrcc} \hline 
Cluster & $\rm Age_{CG97}$(Gyr) & $\rm [Fe/H]_{CG97}$ & $\rm Age_{ZW84}(Gyr)$ & $\rm [Fe/H]_{ZW84}$ & $\rm 
R_{gc}$(Kpc) & $\rm HB_{type}$ & phot & source\footnote{ 
\citet[ T01]{tca:01}, \citet[ H97]{hbv:97}, 
\citet[ SDC91]{sdc:91}, \citet[ H99]{hpc:99}, 
\citet[ SF95]{sf:95}, 
\citet[ H00]{hjz:00}, \citet[ C00]{csa:00},  
\citet[ P01]{pfp:01}, \citet[ P99]{pzk:99}, \citet[ RFV88]{rfv:88}, 
\citet[ CRF91]{crf:91}, \citet[ B98]{bcpf:98}, 
\citet[ S99]{sbh:99}, \citet[ SM86]{sm:86}.}\\ 
\hline 
{\bf NGC~104}  & 10.7 $\pm$ 1.0& -0.78& 10.7 $\pm$ 1.0&-0.71 & 7.4 & -0.99 & $VI$ & R99\\ 
NGC~288  & 11.3 $\pm$ 1.1& -1.14& 11.9 $\pm$ 1.1&-1.40 & 11.6 & 0.98 & $VI$ & R99\\ 
NGC~362  &  8.7 $\pm$ 1.5& -1.09&9.5 $\pm$ 1.5 &-1.33 & 9.3 & -0.87 & $VI$ & R99\\ 
NGC~1261 &  8.6 $\pm$ 1.1& -1.08& 9.1 $\pm$ 1.1 &-1.32 & 18.2 & -0.71 & $VI$ & R99\\ 
NGC~1851 &  9.2 $\pm$ 1.1& -1.03&9.1 $\pm$ 1.1 &-1.23 & 16.7 & -0.36 & $VI$ & R99\\ 
NGC~1904 & 11.7 $\pm$ 1.3& -1.37&12.6 $\pm$ 1.3 &-1.67 & 18.8 & 0.89 & $VI$ & R99\\ 
NGC~2298 & 12.6  $\pm$ 1.4& -1.71&  12.9 $\pm$ 1.4 &-1.85 & 15.7 & 0.93 & $VI$ & T01\\ 
NGC~2419 & 12.3  $\pm$ 1.0& -2.14& 12.8 $\pm$ 1.0 &-2.10 & 91.5 & 0.86 & $VI$ & H97\\ 
NGC~2808 &  9.3  $\pm$ 1.1& -1.11&10.2  $\pm$ 1.1 &-1.36 & 11.0 & -0.49 & $VI$ & R99\\ 
NGC~3201 & 11.3  $\pm$ 1.1& -1.24&12.1  $\pm$ 1.1 &-1.53 &  9.0 & 0.08 & $VI$ & R99\\ 
NGC~4590 & 11.2  $\pm$ 0.9& -2.00&11.2  $\pm$ 0.9 &-2.11 & 10.1 & 0.17 & $VI$ & R99\\ 
NGC~5053 & 10.8  $\pm$ 0.9& -1.98&10.8  $\pm$ 0.9 &-2.10 & 16.9 & 0.52 & $VI$ & R99\\ 
{\bf NGC~5272} & 11.3  $\pm$ 0.7& -1.33& 12.1  $\pm$ 0.7 &-1.66 & 12.2 & 0.08 & $VI$ & R99\\ 
NGC~5466 & 12.2  $\pm$ 0.9& -2.13&12.5  $\pm$ 0.9 &-2.22 & 17.2 & 0.58 & $VI$ & R99\\ 
NGC~5897 & 12.3  $\pm$ 1.2& -1.73& 12.4  $\pm$ 1.2&-1.93 & 7.7 & 0.86 & $VI$ & R99\\ 
NGC~5904 & 10.9  $\pm$ 1.1& -1.12&11.6  $\pm$ 1.1 &-1.38 & 6.2 & 0.31 & $VI$ & R99\\ 
NGC~6093 & 12.4  $\pm$ 1.1& -1.47&12.9  $\pm$ 1.1 &-1.75 & 3.8 & 0.93 & $VI$ & R99\\ 
NGC~6101 & 10.7  $\pm$ 1.4& -1.76& 11.0 $\pm$ 1.4 &-1.81 & 11.1 & 0.84 & $BV$ & SDC91\\ 
NGC~6121 & 11.7  $\pm$ 1.1& -1.05&11.9  $\pm$ 1.1 &-1.27 & 5.9 & -0.06 & $VI$ & R99\\ 
{\bf NGC~6171} & 11.7  $\pm$ 0.8& -0.95&11.7  $\pm$ 0.8 &-1.09 & 3.3 & -0.73 & $VI$ & R99\\ 
NGC~6205 & 11.9  $\pm$ 1.1& -1.33& 13.0  $\pm$ 1.3 &-1.63 & 8.7 & 0.97 & $VI$ & R99\\ 
NGC~6218 & 12.5  $\pm$ 1.3& -1.14&12.7  $\pm$ 1.3 &-1.40 & 4.5 & 0.97 & $VI$ & R99\\ 
NGC~6254 & 11.8  $\pm$ 1.1& -1.25& 12.2  $\pm$ 1.1 &-1.55 & 4.6 & 0.98 & $VI$ & R99\\ 
NGC~6341 & 12.3  $\pm$ 0.9& -2.10& 12.8  $\pm$ 0.9 &-2.24 & 9.6 & 0.91 & $VI$ & R99\\ 
NGC~6352 &  9.9  $\pm$ 1.4& -0.70& 9.7  $\pm$ 1.4 &-0.50 & 3.3 & -1.00 & $VI$ & R99\\ 
NGC~6362 & 11.0  $\pm$ 1.3& -0.99&11.1  $\pm$ 1.3 &-1.08 & 5.3 & -0.58 & $VI$ & R99\\ 
NGC~6366 &  9.6  $\pm$ 1.4& -0.73& 9.4  $\pm$ 1.4&-0.58 & 5.0 & -0.97 & $VI$ & R99\\ 
NGC~6397 & 12.1  $\pm$ 1.1& -1.76& 12.5  $\pm$ 1.1 &-1.94 & 6.0 & 0.98 & $VI$ & R99\\ 
NGC~6426 & 12.9  $\pm$ 1.0& -2.11& 13.0 $\pm$ 1.0 &-2.20 & 14.2 & 0.53 & $VI$ & H99\\ 
NGC~6535 & 12.8  $\pm$ 1.2& -1.51&13.1  $\pm$ 1.2 &-1.78 & 3.9 & 1.00 & $VI$ & R99\\ 
NGC~6584 & 11.3  $\pm$ 1.4& -1.30& 12.1 $\pm$ 1.6 &-1.54 & 7.0 & -0.15 & $BV$ & SF95\\ 
NGC~6624 & 10.6  $\pm$ 1.4& -0.70& 10.6  $\pm$ 1.4 &-0.50 & 1.2 & -1.00 & $VI$ & H00\\ 
NGC~6637 & 10.6  $\pm$ 1.4& -0.78& 10.6 $\pm$ 1.4 &-0.72 & 1.6 & -1.00 & $VI$ & H00\\ 
NGC~6652 & 11.4  $\pm$ 1.0& -0.81& 11.4 $\pm$ 1.0 &-0.89 & 2.4 & -1.00 & $VI$ & C00\\ 
NGC~6656 & 12.3  $\pm$ 1.2& -1.41&12.5  $\pm$ 1.2 &-1.75 & 4.9 & 0.91 & $VI$ & R99\\ 
NGC~6681 & 11.5  $\pm$ 1.4& -1.35& 11.9  $\pm$ 1.4&-1.51 & 2.1 & 0.96 & $VI$ & R99\\ 
NGC~6712 & 10.4  $\pm$ 1.4& -0.94& 10.5 $\pm$ 1.4 &-1.07 & 3.5 & -0.64 & $BV$ & P01\\ 
NGC~6723 & 11.6  $\pm$ 1.3& -0.96& 11.6  $\pm$ 1.3 &-1.12 & 2.6 & -0.08 & $VI$ & R99\\ 
NGC~6752 & 12.2  $\pm$ 1.1& -1.24&12.7  $\pm$ 1.1 &-1.54 & 5.2 & 1.00 & $VI$ & R99\\ 
NGC~6779 & 12.3  $\pm$ 1.4& -1.61&12.8  $\pm$ 1.4 &-1.94 & 9.7 & 0.98 & $VI$ & R99\\ 
NGC~6809 & 12.3  $\pm$ 1.7& -1.54&12.4  $\pm$ 1.7 &-1.80 & 3.8 & 0.87 & $VI$ & R99\\ 
NGC~6838 & 10.2  $\pm$ 1.4& -0.73& 10.1  $\pm$ 1.4&-0.58 & 6.7 & -1.00 & $VI$ & R99\\ 
NGC~6934 &  9.6  $\pm$ 1.5& -1.30& 10.0 $\pm$ 1.6 &-1.54 & 14.3 & 0.25 & $BV$ & P99\\ 
{\bf NGC~7078} & 11.7  $\pm$ 0.8& -2.02&11.8  $\pm$ 0.8 &-2.13 & 10.4 & 0.67 & $VI$ & R99\\ 
NGC~7099 & 11.9  $\pm$ 1.4& -1.92& 12.3 $\pm$ 1.4 &-2.05 & 7.1 & 0.89 & $BV$ & RFV88\\ 
NGC~7492 & 12.0  $\pm$ 1.4& -1.41& 12.1 $\pm$ 1.4&-1.51 & 24.9 & 0.81 & $BV$ & CRF91\\ 
Arp~2    & 11.3  $\pm$ 1.4& -1.45& 11.5  $\pm$ 1.4 &-1.84 & 21.4 & 0.86 & $BV$ & B98\\ 
Eridanus &  8.9  $\pm$ 1.6& -1.20&  8.4  $\pm$ 1.6&-1.48 & 95.2 & -1.00 & $VI$ & S99\\ 
IC~4499  & 12.1  $\pm$ 1.4& -1.26& 11.2  $\pm$ 1.2&-1.50 & 15.7 & 0.11 & $BV$ & B98\\ 
Pal~3    &  9.7 $\pm$ 1.3& -1.39& 9.2 $\pm$ 1.3 &-1.57 & 95.9 & -0.50 & $VI$ & S99\\ 
Pal~4    &  9.5  $\pm$ 1.6& -1.07& 9.2 $\pm$ 1.6&-1.58 & 111.8 & -1.00 & $VI$ & S99\\ 
Pal~5    &  9.8  $\pm$ 1.4& -1.24& 10.0 $\pm$ 1.4 &-1.47 & 18.6 & -0.40 & $BV$ & SM86\\ 
Pal~12   &  6.4  $\pm$ 0.9& -0.83& 6.4  $\pm$ 0.9 &-0.82 & 15.9 & -1.00 & $VI$ & R99\\ 
Rup~106  & 10.2  $\pm$ 1.4& -1.49& 10.4  $\pm$ 1.4 &-1.90 & 18.5 & -0.82 & $BV$ & B98\\ 
Terzan~7 &  7.4  $\pm$ 1.4& -0.56&  7.5  $\pm$ 1.4 &-1.00 & 16.0 & -1.00 & $BV$ & B98\\ 
\hline 
\label{data} 
\end{tabular} 
\vspace*{-0.6cm} 
\end{minipage} 
\end{table*} 
 
The final results for the ages of the entire sample are displayed in  
Table~\ref{data}. In case of non-template clusters, the final error in 
the absolute age has been derived by adding in quadrature the error in the age 
of the corresponding template cluster (which is determined by the 
errors in $\Delta V$ and the internal errors in the adopted metallicity scale) 
to the error in the relative age estimate (which is determined by the 
errors in the differences in $\Delta (V-I)$ or $\Delta (B-V)$, 
and the internal errors in the adopted metallicity scale). 
 
\begin{figure} 
\psfig{figure=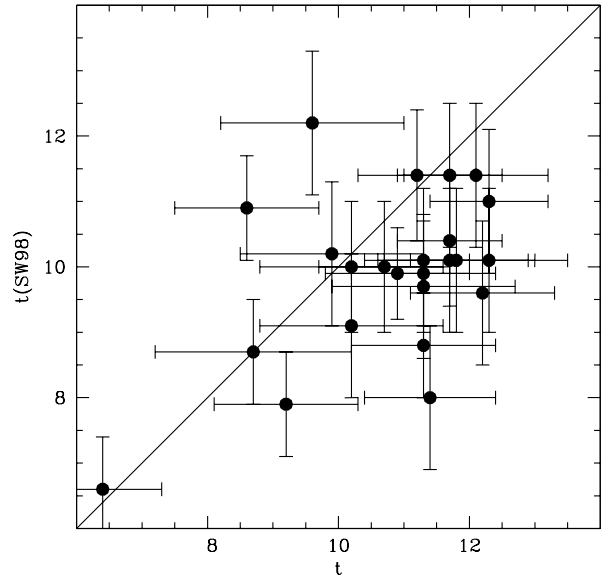,width=8.3cm,clip=} 
\caption[]{Comparison of the absolute ages (in Gyr) obtained by SW98 
(vertical axis) with the values found  
in the present work (horizontal axis), in case of 24 common clusters  
for which we have used new photometric data. 
The solid line represents the locus of equal ages.} 
\label{compage} 
\end{figure}

There are 24 clusters in common with the SW98 sample, for which we have 
now employed more recent and more homogeneous photometry
(mainly in the $VI$ bands); 21 of these 
clusters belong to the R99 sample.  
In Fig.~\ref{compage} we compare the absolute ages derived by 
SW98 (where essentially the same [Fe/H] scale was employed) and 
the ones we now obtained for these 24 clusters in common. There is a 
clear average shift towards higher values in our new sample, which is  
due to the new photometric data; the 
average difference is of the order of 1 Gyr. 
However, in case of two 
metal-rich clusters, namely NGC~6366 and NGC~1261 (new $VI$  
photometries from R99), we get ages considerably lower by $\sim$2.5 
Gyr.

\begin{figure} 
\psfig{figure=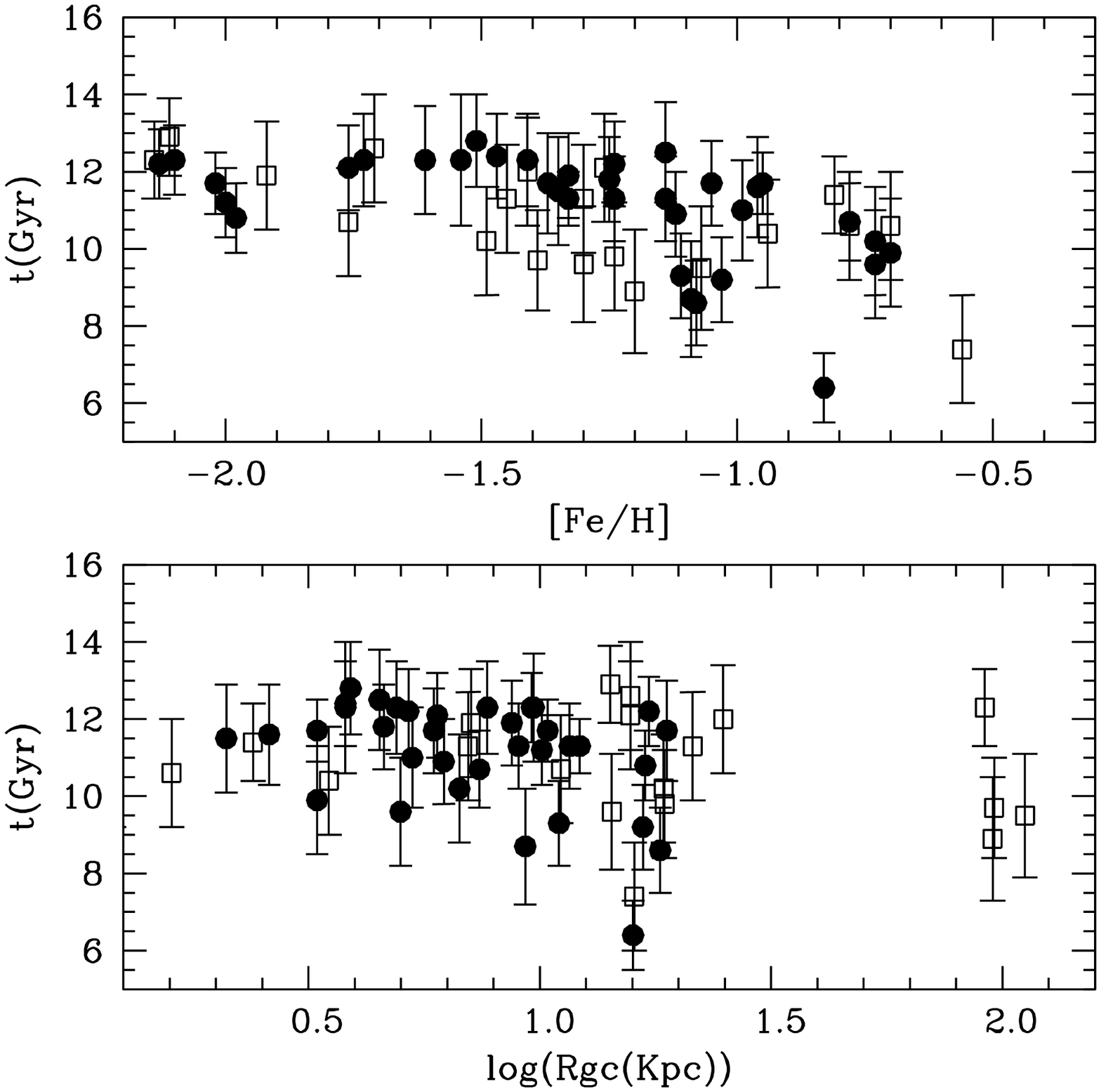,width=8.3cm,clip=} 
\caption[]{Distribution of the absolute ages (in Gyr) as 
function of [Fe/H] (upper panel) and $\rm R_{gc}$ (lower panel). 
Filled circles correspond to clusters in the R99 sample, 
while open squares denote the remaining ones.} 
\label{ages} 
\end{figure}

Fig.~\ref{ages} displays the absolute ages as 
a function of [Fe/H] and galactocentric distance ($\rm 
R_{gc}$) for our whole sample, as reported in  
Table~\ref{data}. Filled circles correspond to clusters in the R99 sample, 
while open squares denote the remaining ones. There are no significant 
differences between these two sub-samples (see also Sect.~4). 
In case of the clusters' $\rm R_{gc}$ we have adopted, as a reference, 
the values provided by \citet{h:96}. Strictly speaking, $\rm R_{gc}$ 
depends on the cluster distance modulus, but for differences of the 
order of 0.1--0.15~mag around the \citet{h:96} values, $\rm R_{gc}$ is 
not changed appreciably, and in case of, e.g.,  
the 35 clusters with ZAHB levels from R99, we determined 
distance moduli within this range from the \citet{h:96} values. 
 
\begin{figure} 
\psfig{figure=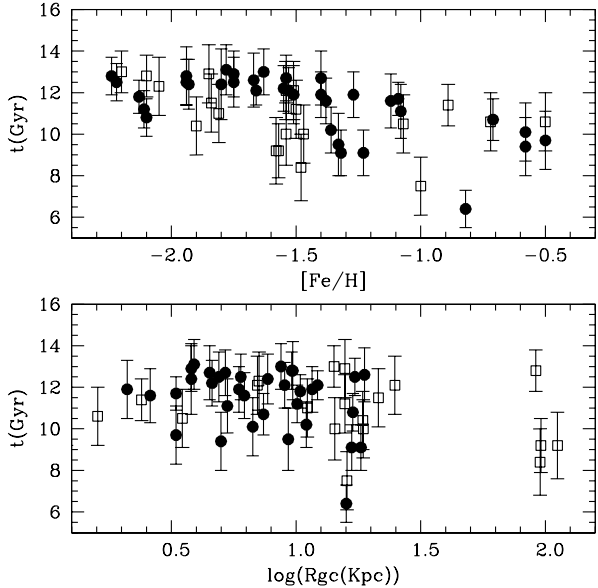,width=8.3cm,clip=} 
\caption[]{ As in Fig.~\ref{ages}, but assuming the ZW84 [Fe/H] scale 
for the clusters instead of the CG97 one in the previous figure.} 
\label{ages2} 
\end{figure}

It is interesting to determine the age difference for two  
classic second-parameter pairs, namely NGC~288--NGC~362  
and M~3--M~13 (NGC~6205), both contained in our sample. 
When deriving errors in age differences within the same metallicity group,  
we have just to take  into account the error 
coming from the use of $\Delta (V-I)$ as the relative age indicator, 
without adding the contribution of the error in the absolute age of 
the template cluster (as is done in Table~\ref{data}),  
because for typical absolute age errors of $\pm$1 Gyr, 
the differential properties of $\Delta (V-I)$ as a function of 
metallicity and age are negligibly affected. 
 
For the pair NGC~288-NGC~362 we obtain an age difference  
of 2.6$\pm$1.5 Gyr (NGC~288 being older),  
in good agreement with the results by \citet{b:01}, but in contrast to 
SW97, where the two clusters appeared to be coeval. The reason for 
this change lies in the new photometric data. 
According to the simulations by \citet{cbl:2001}, 
this age difference, coupled with the absolute ages we obtain and the 
use of the CG97 [Fe/H] scale, can explain the overall different HB morphology; 
however, canonical HB models appear unable to reproduce  
the detailed morphology of the red end of NGC~288 HB \citep{cbl:2001}. 
 
In case of M~3-M~13 we obtain a difference of 0.6$\pm$1.0 Gyr (M13 older),  
much less significant than the 1.7$\pm$0.7 Gyr as obtained 
by \citet{ryl:01} from their $BV$ photometry. M~13 was not contained in 
our previous work. The same negligible age difference we obtain from 
the $\Delta V$ values derived by R99; it seems therefore that the 
discrepancy between our and \citet{ryl:01} result is due to real 
differences in the photometric data, and not to the use of 
different passbands for the TO-RGB colour differences and 
inconsistency in the colours of the theoretical isochrones. Problems with the  
calibration of the photometry may possibly lead to this kind of discrepancy.  
\citet{ryl:01} noticed, for example, that their derived fiducial line 
for M~13 agrees well with those obtained by  
\citet{rf:86} and \citet{ybs:2000}, but differs in the main sequence 
and subgiant branch region from the fiducial by \citet{pff:98}. 
 
It is also interesting to notice that 
Arp~2 and Rup~106 do not appear much younger than the bulk of the clusters 
at their metallicity; the reason why \citet{bcpf:98} found higher age 
differences is mainly the fact that their adopted
clusters' absolute ages are higher. 
As discussed in detail in, e.g., \citet{psw:98}, lower absolute ages 
imply smaller age differences for a given observed distribution of 
$\Delta V$, $\Delta(B-V)$ or $\Delta(V-I)$ values.
As discussed in SW98 there are preliminary 
indications that Rup~106 and also Pal~12 may not show $\alpha$-element 
enhancement. In this case, their ages displayed 
in Table~\ref{data} should be increased by about 1~Gyr.  
 
To highlight the effect of the present uncertainties in the [Fe/H] scale, 
we have also derived ages, as a test, by using the 
[Fe/H] values given by \citet{rhs:97} 
on the ZW84 scale (internal accuracy again of the order of 0.10 dex), 
complemented, if needed, by data in 
the original ZW84 paper or coming from other spectral 
indices calibrated on the ZW84 scale. In this case 
the [Fe/H] range spanned by our sample is larger than  
the CG97 scale. We have therefore divided the sample into 5 groups, 
having as template clusters  
M~15 ($-2.3 \leq $[Fe/H]$ \leq -$2.0), NGC~6656 ($-1.99 \leq $[Fe/H]$ \leq -$1.7), 
M~3 ($-1.69 \leq $[Fe/H]$ \leq -$1.4), NGC~6171 ($-1.39 \leq $[Fe/H]$ \leq 
-$0.9) and 47~Tuc ($-0.89 \leq $[Fe/H]$ \leq -$0.5). 
Fig.~\ref{ages2} shows the age distribution resulting from the 
AM-method as a function 
of [Fe/H] and $\rm R_{gc}$, when using the ZW84 [Fe/H] scale. 
Using this metallicity scale, the age differences between the two second 
parameter pairs discussed remain basically unchanged. In fact, in 
case of NGC~288--NGC~362 we obtain now a difference of   
2.4$\pm$1.5 Gyr, and for M~3--M~13 we get 0.9$\pm$1.0 Gyr. 
A detailed analysis of the age distributions displayed in Fig.~\ref{ages} 
and Fig.~\ref{ages2} will be presented in the last section. 
 
\section{Colour-based relative ages} 
 
We consider it important that isochrones -- together with colour 
transformations -- yield ages largely independent of the age 
indicator. If consistency is achieved, one could then in principle 
apply different age indicators to different clusters, 
according to the quality of the available data, and still have   
consistent ages for all the sample. 
 
In the last section we showed that the AM-method yields results which 
are consistent with absolute and 
relative ages from the $\Delta 
V$ -- or vertical method. Another age indicator -- the horizontal one -- 
is the colour  
difference between TO and RGB, which we used already 
in the AM-method to obtain relative ages within each metallicity 
bin. In the following, we will call this quantity simply $\delta C$, 
independent of the actual colour it refers to. From 
an observational point of view this age-indicator is the 
easiest one to determine, since it uses 
a CMD portion which is common to all  
clusters, is well-populated, and thus -- for sufficiently deep photometry -- 
observable with high accuracy. Reddening is not important, as long as 
it is low and not differential. However, the difficulties start when 
using it for absolute ages or over a wide range of metallicities; in 
this case, observed individual $\delta C$ values have to be compared to 
theoretically predicted ones,  
and colour transformations -- with their known  
uncertainties -- must be able to predict with extreme accuracy 
the absolute values of colours over a  
large [Fe/H] and effective temperature range. 
 
\begin{figure} 
\psfig{figure=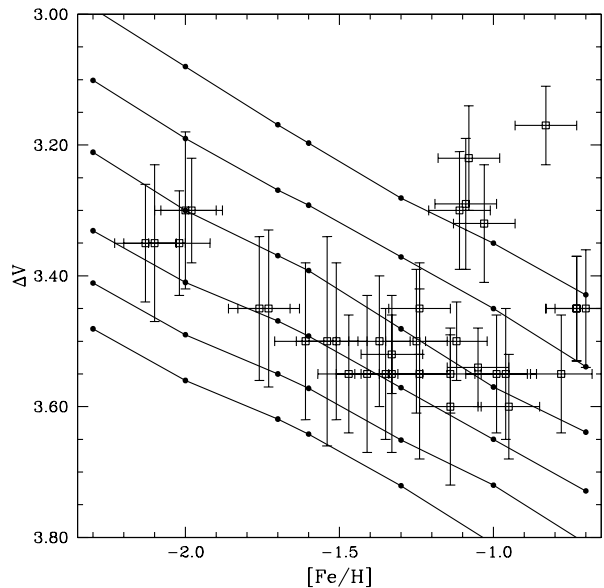,width=8.3cm,clip=} 
\caption[]{Observed $\Delta V$ values for the 35 clusters from R99 
along with the theoretical values from our 
isochrones, plotted for 14--9~Gyrs (bottom to top in steps of 
-1~Gyr).} 
\label{dv} 
\end{figure}

Because of the easy observational accessibility, \cite{bcpf:98} 
developed a semi-empirical relative-age determination method using 
$\delta C$ in the $BV$ plane. It sets  
out by selecting a group of coeval clusters of all metallicities, 
which defines the empirical function $\delta C(\mathrm{[Fe/H]})$ at 
the age $t_0$ of this group. For this step,  
$\Delta V$ and theoretical isochrones are needed. It turns out that 
membership to the coeval group is rather independent of the isochrones 
used, but not the actual value for $t_0$.  
The result for our own isochrones is shown in Fig.~\ref{dv}. 
The empirical $\delta C(\mathrm{[Fe/H]})$-relation 
at $t_0$ can be 
compared to the theoretical one. Agreement is not necessarily given 
\citep{bcpf:98} and depends on isochrone set and colour transformation. 
 
In the next step clusters of ages   
differing from $t_0$ are used to derive the relation $\delta C(t)$ at 
various $\mathrm{[Fe/H]}$ points. Age differences are again determined 
from the vertical method and isochrones. Combined with the first step, 
one obtains $\delta C$ as a function of age and metallicity, such that 
the observed values for $\delta C$ along with $\mathrm{[Fe/H]}$ yield 
the age straightforwardly.  
This makes the method very attractive for large cluster samples.  
Because of the problems mentioned above the method  
works best for relative ages, but one has to keep in mind that age 
differences also depend on the absolute age (SW97). In 
\cite{psw:98} we showed that our own isochrones and colour 
transformations to $(B-V)$ result in a $\delta C(\mathrm{[Fe/H]})$ 
relation consistent with observations and that  
relative ages from this method agree well with those we obtain from 
the AM-method, which might be understood as a ``piecewise'' 
application of that of \cite{bcpf:98}.  
 
This same idea has then been used by R99 for their clusters 
observed in $V$ and $I$ \citep{rasp:2000,rpsa:2000}.  
They employed isochrones from \citet{scl:97}  
and a pre-release of \citet{vsria:2000} models 
for the theoretical determination of  
$\delta C(t)$ at various $\mathrm{[Fe/H]}$, and the TO brightness 
at various ages and [Fe/H]; as for the HB brightness, they considered 
the empirical ZAHB brightness relationship by \citet{cgc:2000}, lacking 
access to ZAHB models consistent with the isochrones adopted. 
 
As in \cite{bcpf:98} the comparison of ages from $\Delta V$ and 
$\Delta(V-I)$ revealed inconsistencies:  
the absolute age $t_0$ of 
the coeval reference sample depends on age indicator and 
isochrones: it is (ages in Gyr) $\sim$14 for \cite{vsria:2000} 
and $\sim$15 for \cite{scl:97}, according to $\Delta V$, 
but, respectively, 16 and 13 for 
$\Delta(V-I)$, i.e.\ there are changes of 2~Gyr with respect to 
the $\Delta V$ ages, but they are 
in opposite directions for the two isochrone sets.

\begin{figure} 
\psfig{figure=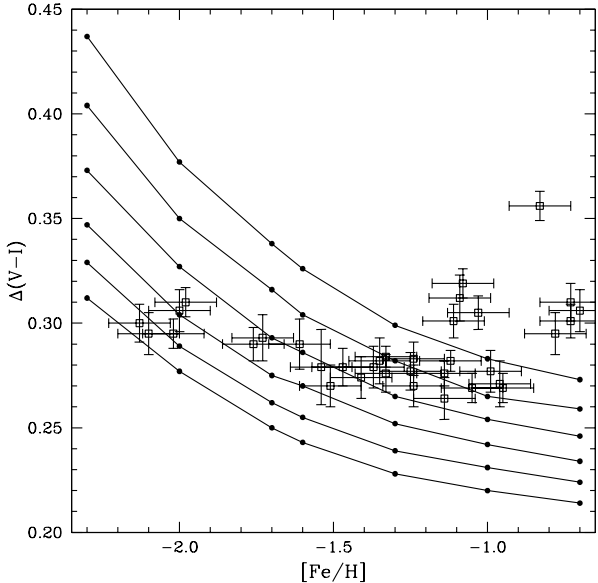,width=8.3cm,clip=} 
\caption[]{The observed horizontal relative age indicator $\Delta(V-I)$ 
for the 35 clusters from R99 along with the theoretical values from our 
isochrones, plotted for 14--9~Gyrs (bottom to top in steps of 
-1~Gyr). Isochrones were transformed to $(V-I)$ 
using the colour transformation selected in \cite{ws:99}.} 
\label{dvi} 
\end{figure} 
 
\begin{figure} 
\psfig{figure=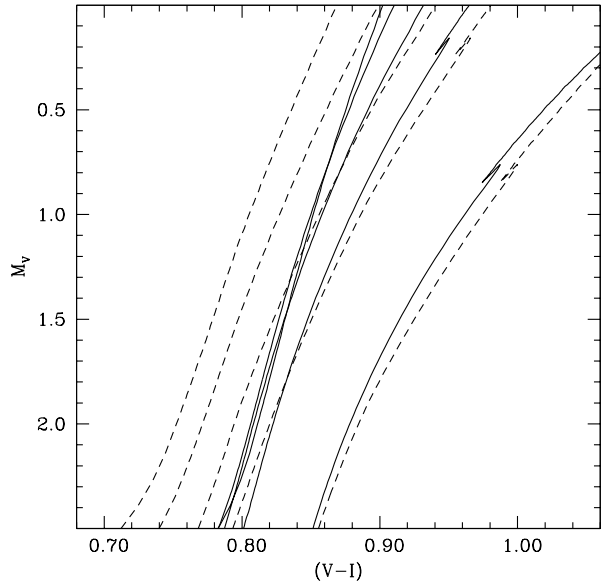,width=8.3cm,clip=}  
\caption[]{Comparison of RGB $(V-I)$ colours for isochrones of 12~Gyr 
and [Fe/H]=$-2.3, -2.0, -1.6, -1.3, -0.7$ 
(from left to right). Solid lines 
correspond to the colour transformation selected in \cite{ws:99}, dashed ones 
to those by \cite{aam:99}.} 
\label{isovi} 
\end{figure} 
 
We already demonstrated (Fig.~\ref{compage2}) that the relative AM-ages, 
which are basically obtained from the horizontal method, 
agree well with the $\Delta V$ values for the R99 sample. 
In particular,  
the majority of clusters is coeval at $\approx 12$~Gyr with a small 
metal-rich group at ages lower by 1--3~Gyr (see Fig.~\ref{dv}). 
According to the (global) horizontal age indicator $\Delta(V-I)$, 
however, there is a  
clear age increase of almost 3~Gyr within the coeval 
group (Fig.~\ref{dvi}). This effect is most pronounced at the lowest 
metallicities. The $\Delta(V-I)$ therefore, when used as 
an age indicator across all the metallicity range, 
along with our isochrones 
and colour transformations, produces inconsistent results with respect 
to $\Delta V$. 
Note, however, that the horizontal average age of the ``coeval'' group is around 
11~Gyr and -- within the errors -- consistent with that of the 
vertical one (Fig.~\ref{dv}), and that the whole  
discrepancy in the $\Delta(V-I)$ relative ages is due to a 
variation in $\Delta(V-I)$ of only $\approx 0.05$~mag. 
This is a strong indication that theoretical colours have to be very 
accurate for the whole metallicity range of globular clusters in order 
that this method can return reliable results. 
 
\begin{figure} 
\psfig{figure=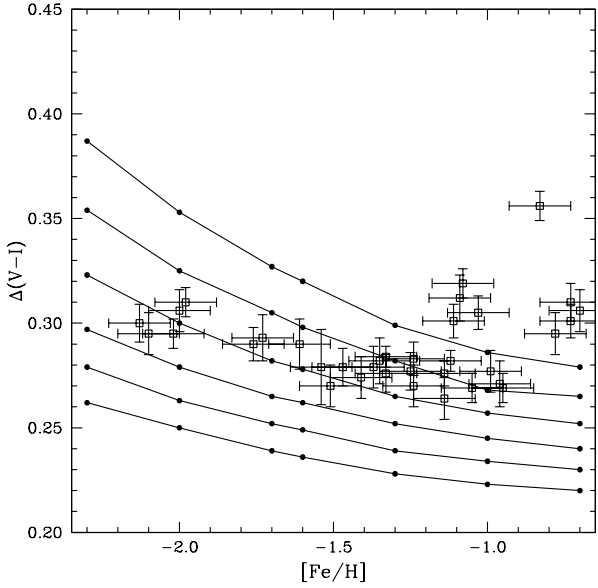,width=8.3cm,clip=} 
\caption[]{As Fig.~\ref{dvi}, but with the colour 
transformations by \cite{aam:99}.} 
\label{dvi2} 
\end{figure}

To emphasize this point we show in Fig.~\ref{isovi} the 
predicted RGB colours for 12~Gyr-isochrones of different metallicities  
as obtained by using the transformations of \cite{ws:99} 
or \cite{aam:99}. Evidently, in the former case, which is 
based on \cite{vb:98}, the dependence of colour on metallicity is 
non-monotonic and RGBs are piling up at a certain bluest level, while 
for \cite{aam:99} the behaviour is much more regular. Taking this as the 
``true'' behaviour, \cite{vb:98} RGBs would be too red for the lowest 
metallicities by about 0.04-0.06~mag, i.e.\ ages would be too high by the 
corresponding age difference of 2-3~Gyr, just as visible in 
Fig.~\ref{dvi}. Note that at higher metallicities the two 
transformations agree well. 
Repeating therefore the horizontal absolute age determination, 
but using the transformation by \cite{aam:99}, we obtain 
the results displayed in 
Fig.~\ref{dvi2}. While there is still a small age gradient across the 
main group of clusters, the total age difference is reduced to within 
1.5~Gyr, and is consistent within the errors  
with the results of the AM-method --  which, as already discussed in the 
previous section, are unaffected by the colour transformation choice 
-- and from the $\Delta V$. This shows 
that our isochrones together with the 
colour transformations by \cite{aam:99} for the RGB are able to 
produce stable relative ages for the whole metallicity range of GCs 
independent of the use of the AM-method, the $\Delta V$, or $\delta 
C$ age indicator.

 
\section{Discussion} 
 
Having fully assessed the consistency of the age distribution  
within our cluster 
sample, the results displayed in Fig.~\ref{ages} can be used to investigate 
the formation mechanism of the Galaxy. In the 
following we update the discussion in SW97 and SW98, taking into account 
this much larger cluster sample.  
 
There are about 150 globular clusters known in the Galaxy 
(\citealt{h:96}), and probably the total population is of 
at most about 200 objects (\citealt{h:76}); therefore, our sample 
contains between 25\% and 35\% of the Galactic globular  
cluster population, and should provide statistically more significant   
data about the Galaxy formation mechanism, with respect 
to previous works. 
 
Two fundamental pieces of information need 
to be extracted from our age-analysis, namely the age distribution  
of the clusters with respect to their initial chemical 
composition (whose diagnostic is [Fe/H]), and with respect to their 
position within the Galaxy (whose diagnostic is  
$\rm R_{gc}$). 
To detect an intrinsic age spread -- that is, not due to the error 
bar on the age determinations -- among a given sample of objects 
which shows an age dispersion $\sigma_\mathrm{obs}$, 
we have followed the same procedure as in SW97  
(see also \citealt{cds:96}). 
More in detail,  
we calculate an "expected" distribution for the assumption of no  
intrinsic age range ($\sigma_\mathrm{range}$=0) by randomly 
generating, for each object, 10000 ages using a Gaussian distribution.  
The mean of the distribution is given by the mean age of the clusters 
in the given sample, and the $\sigma_\mathrm{exp}$ by the typical  
error in the individual age determinations  
(in this case we consider the total error given in  
Table~\ref{data}, since we compare ages across multiple metallicity ranges).  
This is repeated for all clusters in the sample, and the 
F-test is then applied to the observed age distribution  
and this synthetic one, to test if they both show the same 
variance. 
We state that an age range exists if the probability that the two 
distributions have the same variance is smaller than 5\%. If this is  
true, the intrinsic age dispersion is then estimated by 
$\sigma_\mathrm{range}$=($\sigma_\mathrm{obs}^2 - \sigma_\mathrm{exp}^2)^{0.5}$. 
 
\subsection{Age-metallicity relationship} 
 
When considering the complete sample of 55 objects, we obtain that 
all clusters with [Fe/H]$\leq -$1.2 result to be coeval 
(within the stated error bars), with an average age  
$\langle$t$\rangle$=11.7$\pm$0.7 Gyr, where the dispersion of 0.7 Gyr 
is entirely due to the errors in the age determination. In case of 
[Fe/H]$\geq -$1.2 a statistically significant age spread is 
present. The average age is $\langle$t$\rangle$=10.1 Gyr, 
with an intrinsic spread $\sigma_\mathrm{range}$=1.9 Gyr.  
In this same [Fe/H] range we have detected, overimposed to the age 
spread, a marginally significant age gradient  
$\delta t/\delta \rm[Fe/H]$=$-2.4\pm$1.8 Gyr/dex. 
 
Our complete sample contains 3 clusters generally considered to be 
part of the thick disk of the Galaxy (NGC~104, NGC~6352, NGC~6838), and 
2 clusters possibly belonging to the bulge population (NGC~6637, NGC~6324). 
All of these five clusters are metal rich and located in the inner 
halo. If we repeat our previous analysis excluding these 5 objects, and 
therefore considering only bona fide halo clusters, we obtain an 
average age $\langle$t$\rangle$=10.0 Gyr for the metal rich group, 
with $\sigma_\mathrm{range}$=2.5 Gyr, and a more significant 
$\delta t/\delta \rm[Fe/H]$=$-4.3\pm$2.4 Gyr/dex.

\subsection{Age-galactocentric distance relationship} 
 
The clusters in the so-called inner halo ($\rm R_{gc}\leq 8$ kpc) appear to be 
coeval, with $\langle$t$\rangle$=11.4$\pm$0.8 Gyr, while the outer 
halo clusters display an intrinsic age range $\sigma_\mathrm{range}$=2.5 Gyr. 
The average age of the outer halo sample is $\langle$t$\rangle$=10.7 
Gyr, lower than the inner halo one, simply due to the age spread 
towards lower ages, but there is no significant age gradient in the 
outer halo. This would imply that globular clusters  
have started forming at the same 
time throughout the Galaxy -- this holds of course in case the actual 
$\rm R_{gc}$ is a good approximation of the value at the beginning of 
Galactic formation; see, e.g., the discussion by \citet{sz:78} 
on this subject -- with younger and more metal rich  
objects being formed or possibly accreted  
only at large $\rm R_{gc}$ values. 
This scenario is very similar to the one proposed by 
\citet{sz:78}, where the inner halo collapsed on short timescales, 
while the outer halo underwent a much longer formation period,  
accreting material over several Gyr. 
 
When considering only bona fide halo clusters the inner halo 
appears still to be coeval, its average age being basically 
unchanged ($\langle$t$\rangle$=11.7$\pm$0.6 Gyr).

\subsection{Influence of the sample size and metallicity scale} 
 
The previously described results about the age distribution with 
respect to [Fe/H] and $\rm R_{gc}$ are unchanged  
if we restrict our analysis to the 
more homogeneous sample of 35 clusters from R99. 
Moreover, the age trends with respect to [Fe/H] and $\rm R_{gc}$  
we derive are similar to the results obtained by R99 with their different 
technique, and statistically  
more significant because of the larger cluster sample.   
When comparing our new results with SW98 we find that  
the general picture of age versus [Fe/H] is not greatly 
modified. In our previous works we obtained an approximate  
constant age for the oldest clusters at all metallicities, plus the onset of an 
age spread above a given [Fe/H]. 
However, while in SW97 and SW98 we did not find any clear correlation 
between age and $\rm R_{gc}$, we find now a difference between the 
age distribution of the inner and outer halo clusters. This is due to 
the much larger cluster sample and also to the upward revision of the 
age of NGC6652, which in SW97 and SW98  resulted to be much 
younger than the other inner halo clusters. 
 
The details of the scenario emerging from our age determination 
are however dependent on the selected [Fe/H] scale. 
When we consider the ages obtained with the ZW84 scale displayed 
in Fig.~\ref{ages2}, we find that only clusters with 
[Fe/H]$\leq -$1.6 are coeval, with an average  
age $\langle$t$\rangle$=12.2$\pm$0.6 Gyr, 
slightly larger than in the case of the CG97 metallicities. For larger 
[Fe/H] values an intrinsic spread  $\sigma_\mathrm{range}$=2.1 Gyr is 
present, with an average age $\langle$t$\rangle$=10.5 Gyr, and  
$\delta t/\delta \rm[Fe/H]$=$-1.3\pm$0.7 Gyr/dex. 
Also the distribution of ages as a function of  
$\rm R_{gc}$ is more homogeneous, with an intrinsic age spread 
appearing only when $\rm R_{gc}$ is larger than $\sim$13 kpc. 
If we restrict the analysis to only bona fide halo clusters, 
we find for [Fe/H]$> -$1.6  
a slope $\delta t/\delta \rm[Fe/H]$=$-2.3\pm$1.1 Gyr/dex,  
the significance of 
which now exceeds the 2$\sigma$ level. 
 
These differences 
come obviously from the different metallicity 
distribution of the clusters. In particular, in the range 
$-1.8 \leq$[Fe/H]$\leq -1.0$, large differences (up to $\sim$0.3 dex) 
exist between the two scales.  
 
Taking into account the effect of the uncertainty in the [Fe/H] scale 
and the cluster sample size, helps 
in explaining discrepant conclusions reached by different 
authors about the existence and the onset of the age spread 
in the cluster population.   
As a test, we have considered the clusters used by  
\citet[V00]{v:00} in his recent analysis; by applying our  
statistical analysis to his age determinations,  
we find the onset  
of an age variation among clusters starting at [Fe/H]$\sim-$1.6, 
lower than the value [Fe/H]=$-$1.2 found by R99, who used the CG97 
scale, and also lower than our results  
when using the CG97 scale. A significant 
age-metallicity relationship is found for [Fe/H]$\ge -$1.6. 
 
V00 derived ages for 26 objects --  
25 of which are also included in our sample -- by employing 
cluster [Fe/H] values close to the ZW84 scale, and  
absolute age determinations based essentially on $\Delta V$  
(more precisely, the distance modulus is determined from fitting the appropriate 
theoretical ZAHB to the observational counterpart, and then the age is 
determined by the isochrone which best fits all the TO region 
of the observed Colour-Magnitude-Diagram). 
 
\begin{figure} 
\psfig{figure=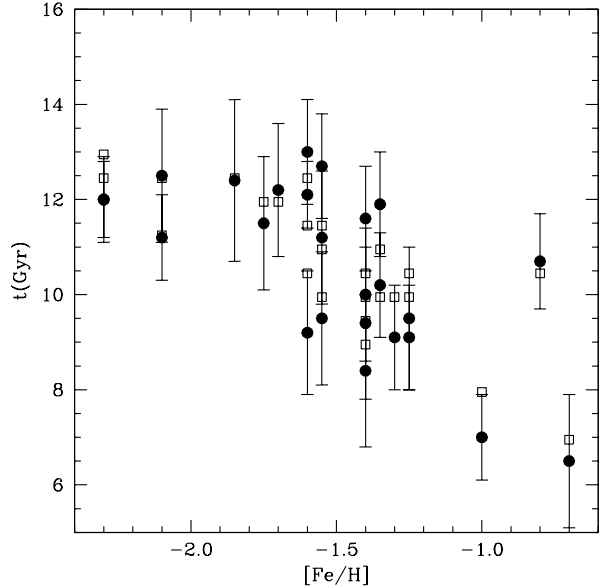,width=8.3cm,clip=} 
\caption[]{Comparison between the ages derived with the AM-method, 
our isochrones and our adopted photometries (filled circles),  
and the ages from V00 (open squares) for 25 clusters in common;  
an offset by $-1$ Gyr has been 
applied to the V00 data. For the sake of clarity V00 error bars (of the 
order of 0.8 Gyr) have been omitted.} 
\label{agevdb} 
\end{figure} 
 
We have rederived the ages for the 25 clusters in common assuming V00 
[Fe/H] values, our isochrones, our method and our adopted 
photometries (which are different 
from the data employed by V00 for the majority of the clusters),  
and the results are displayed in Fig.~\ref{agevdb}, 
together with V00 ages shifted by $-1$ Gyr. 
When this shift of the absolute age is applied --  which reflects mainly the  
ZAHB brightness difference between the different sets of models 
employed --  
the age distribution appears quite similar, all pairs of values being 
in agreement within the respective error bar.  
This consistency is also confirmed by the 
application of a K-S test to the two samples. 
If we restrict the analysis to bona fide halo clusters (thus 
not considering 47~Tuc in the V00 sample), 
for [Fe/H]$\geq -$1.6 we obtain an age-metallicity 
relationship with slope $\delta t/\delta \rm[Fe/H]$=$-4.7\pm$0.8 Gyr/dex, 
in good agreement with the value  
$\delta t/\delta \rm[Fe/H]$=$-5.7\pm$1.2 Gyr/dex 
one obtains from V00 ages. 
This slope is about a factor of two higher than the one we obtained 
with our full sample of bona fide halo clusters 
coupled with the ZW84 [Fe/H] scale. 
 
\subsection{Second parameter and age} 
 
With our large cluster sample it is also possible to investigate the 
global correlation between [Fe/H], HB type and age for the Galactic globular 
cluster system as a whole. We have already discussed  
in Sec.~2 two specific pairs of second-parameter clusters; in case 
of NGC~288-NGC~362 we found that age could be the main 
reason for the different HB types, while we do not find a 
significant age difference for the pair M~3-M~13.  
 
\begin{figure} 
\psfig{figure=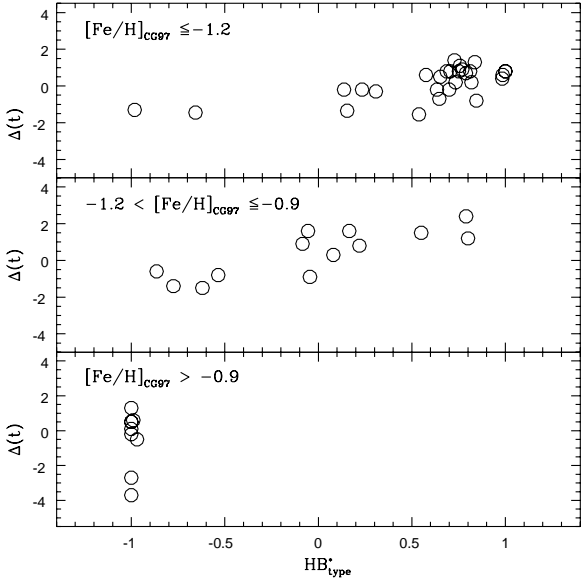,width=8.3cm,clip=} 
\caption[]{Relationship between age differences $\Delta$(t)  
and normalized HB type for clusters in the [Fe/H] ranges 
(on the CG97 scale) indicated in the three panels.  
See text for details.} 
\label{HBtype1} 
\end{figure}

\begin{figure} 
\psfig{figure=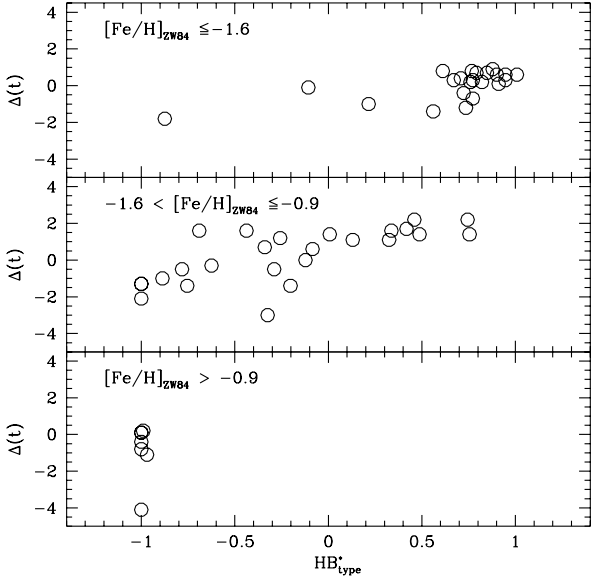,width=8.3cm,clip=} 
\caption[]{As in Fig.~\ref{HBtype1} but considering the ZW84 [Fe/H] 
scale. See text for details} 
\label{HBtype2} 
\end{figure} 
 
In order to detect possible correlations between age and HB type  
(defined as $\rm HB_{type}$=(B-V)/(B+V+R), where B,V,R are the 
number of HB stars 
located, respectively, at the blue side of the instability strip, in 
the instability strip and at its red side) 
superimposed on the relationship between HB type and [Fe/H],    
we have performed the following test.  
We have first considered the samples of coeval clusters with  
$\rm R_{gc}\leq 8$ for the CG97 [Fe/H] scale, and  
$\rm R_{gc}\leq 13$ for the ZW84 one; 
since in both cases they span the entire metallicity range 
covered by the whole sample, we use them to determine the variation of  
$\rm HB_{type}$ with respect to [Fe/H], at constant age. 
The entire sample has then been divided into 
different metallicity bins, specified in  
Fig.~\ref{HBtype1} and Fig.~\ref{HBtype2}. 
For the most metal poor bin 
the average age of the sample has been 
subtracted from the individual cluster ages; in case of the two most
metal rich bins, the average age of their combined 
sample has been subtracted from the individual values.
These normalized individual ages are denoted as $\Delta(t)$. 
To remove the influence of the first parameter ([Fe/H]),  
in addition, a reference [Fe/H] has been selected for each range, 
and the individual values of $\rm HB_{type}$ 
(taken from \citealt{h:96} and displayed in Table~\ref{data}) have been 
corrected using the derivative 
$\Delta \rm HB_{type}$/$\Delta$ [Fe/H] (determined from the previously 
mentioned coeval samples) and considering the difference between 
the actual cluster [Fe/H] value and the reference one. 
By plotting $\Delta(t)$ as function of $\rm HB_{type}^{*}$, that is, 
the observed HB type normalized to the reference [Fe/H] value,  
one should in principle be able to isolate an hypothetical correlation 
between age differences and HB colour differences, independent of the effect 
of [Fe/H] variations among clusters. 
The reference [Fe/H] values are, respectively, [Fe/H]=$-$1.75  
and $-$1.1 for the CG97 scale, and [Fe/H]=$-$2.0 and $-$1.3 for the ZW84 one. 
We did not correct the HB type for the most metal rich cluster groups 
in Fig.~\ref{HBtype1} and Fig.~\ref{HBtype2},  
since the value of $\rm HB_{type}$ saturates and does not provide  
useful information about a possible correlation with age. 
 
The upper panels of both Fig.~\ref{HBtype1} and Fig.~\ref{HBtype2} 
display [Fe/H] ranges where our F-test analysis did not detect a 
significant age spread compared to the errors in the age estimates. 
In these ranges there is some spread in HB type but no significant 
correlation with age is found. 
In the middle panels the F-test analysis detected significant age 
spreads and also a statistically 
significant correlation with HB type is found. The 
bottom panels display clusters showing an age spread but their in this 
[Fe/H] range the  $\rm HB_{type}$ parameter saturates and no 
relevant information can be deduced. 
  
\section{Summary} 
 
We have determined absolute and relative ages for a large sample of 
globular clusters, using new and homogeneous photometric data, mostly in $V$ 
and $I$. We show that different age determination methods result in very 
similar ages, if colour transformations are carefully chosen for those 
methods using colour differences. Our preferred AM-method
appears however insensitive to the colour transformations choice.
In a few cases, new photometric data have 
led to a substantial change in age compared to our previous works; this 
source of uncertainty should not be ignored.
 
From the analysis of our large sample of 55 Galactic globular  
clusters, we have found that: 
 
i) Irrespective of the adopted [Fe/H] scale, the metal poorer 
clusters are coeval within $\sim$1 Gyr, while an age spread 
appears at higher metallicites. The onset of the age spread is at  
[Fe/H]$\sim -$1.2 when using  
[Fe/H] values on the CG97 scale, and [Fe/H]$\sim -$1.6 if one adopts 
the ZW84 scale. 
 
ii) An age-metallicity relationship 
exists among the non-coeval clusters. In case of the CG97 scale we find   
$\delta t/\delta \rm[Fe/H]$=$-2.4\pm$1.8 Gyr/dex when considering the complete 
sample; if we restrict the analysis only to bona fide halo clusters we 
obtain a more significant slope 
$\delta t/\delta \rm[Fe/H]$=$-4.3\pm$2.4 Gyr/dex. 
In case of the ZW84 scale we obtain  
$\delta t/\delta \rm[Fe/H]$=$-1.3\pm$0.7 Gyr/dex for the complete sample, 
and $\delta t/\delta \rm[Fe/H]$=$-2.3\pm$1.1 Gyr/dex for the pure halo sample. 
 
iii) Irrespective of the adopted [Fe/H] scale, clusters closer to 
the galactic centre are coeval, while an age spread appears at larger 
galactocentric distances. In case of the CG97 [Fe/H] scale the age 
spread starts at $\rm R_{gc}$ larger than $\sim$ 8 kpc, while in case 
of the ZW84 scale it starts at $\rm R_{gc}$ above 13 kpc. 
In both cases there is no significant age- $\rm R_{gc}$ relationship 
among clusters showing an age spread.  
 
iv) Our results yield indications that age differences are one main 
factor able to explain the global behaviour of the second 
parameter effect.

\begin{acknowledgements} 
We are grateful to A.~Rosenberg and I.~Saviane for helpful discussions 
and information about the cluster sample and relative ages, and to 
S.~Percival for a preliminary reading of the manuscript.  
We thank and anonymous referee for his/her comments which helped to 
improve the presentation of the results. 
\end{acknowledgements} 
%

\end{document}